\documentclass{elsarticle}

\usepackage[english]{babel}

\usepackage[letterpaper,top=2cm,bottom=2cm,left=3cm,right=3cm,marginparwidth=1.75cm]{geometry}

\usepackage{amsmath,amssymb, mathtools}
\usepackage{graphicx}
\usepackage[colorlinks=true, allcolors=blue]{hyperref}
\usepackage[normalem]{ulem}
\usepackage{bbm}
\usepackage{dsfont}
\mathtoolsset{showonlyrefs}

\def\derc#1#2{\frac{\partial#1}{\partial#2}}
\def\ts{t^{*}}
\def\VT{V^{T}}
\def\Nc{N_c}

\def\lambda{H}

\usepackage{xcolor}

\begin{document}

\begin{frontmatter}
\title{A refractory density approach to a multi-scale SEIRS epidemic model}

\author[inst1]{Anton Chizhov}
\affiliation[inst1]{organization={Centre Inria d'Université Côte d'Azur},
            addressline={2004 Rte des Lucioles}, 
            city={Biot},
            postcode={06410}, 
            country={France}}

\author[inst2]{Laurent Pujo-Menjouet\corref{cor1}}
\ead{pujo@math.univ-lyon1.fr}
\affiliation[inst2]{organization={Universite Claude Bernard Lyon 1, CNRS, Ecole Centrale de Lyon, INSA Lyon, Université Jean Monnet, ICJ, UMR5208, Inria},
            city={Villeurbanne},
            postcode={69622}, 
            country={France}}

\author[inst3,inst4]{Tilo Schwalger}
\affiliation[inst3]{organization={Bernstein Center for Computational Neuroscience},
            addressline={Philippstr. 13, Haus 6}, 
            city={Berlin},
            postcode={10115}, 
            country={Germany}}       
\affiliation[inst4]{organization={Technische Universität Berlin
Institute of Mathematics},
            addressline={Straße des 17. Juni 136}, 
            city={Berlin},
            postcode={10623}, 
            country={Germany}}  

\author[inst1,inst5]{Mattia Sensi}
\affiliation[inst5]{organization={Department of Mathematical Sciences ``G. L. Lagrange'', Politecnico di Torino},
            addressline={Corso Duca degli Abruzzi 24}, 
            city={Torino},
            postcode={10129}, 
            country={Italy}}  
\cortext[cor1]{Corresponding author.}

\date{\today}

\begin{abstract}
We propose a novel multi-scale modeling framework for infectious disease spreading, borrowing ideas and modeling tools from the so-called Refractory Density (RD) approach. We introduce a microscopic model that describes the probability of infection for a single individual and the evolution of the disease within their body. From the individual-level description, we then present the corresponding population-level model of epidemic spreading on the mesoscopic and macroscopic scale. We conclude with numerical illustrations taking into account either a white Gaussian noise or an escape noise to showcase the potential of our approach in producing both transient and asymptotic complex dynamics as well as finite-size fluctuations consistently across multiple scales. A comparison with the epidemiology of coronaviruses is also given to corroborate the qualitative relevance of our new approach.
\end{abstract}

\end{frontmatter}

\section{Introduction}

The susceptibility of an individual to an infectious disease as well as the infectiousness of an infected individual considerably depends on the time of the person's last infection, or ``infection age''. Such dependence on the infection course and history can be modeled by age-structured population models. These models have been used since the seminal paper by Kermack and McKendrick \cite{kermack1927contribution} for a wide variety of mathematical models in epidemics, see \textit{e.g.} \cite{alfaro2019analysis,And88,brauer2009age,d2023optimal,li2019mathematical}. Taking into account the time since the last infection (TSLI) is extremely convenient to introduce a temporal heterogeneity in the population concerning an ongoing epidemic, and to explicitly model a complex disease evolution with consequences both within and between hosts.

Population models structured by this TSLI are deterministic macroscopic models that relate to the idealized limit of an infinitely large population. Two important questions arise here: first, a fundamental question is how such macroscopic models are linked to ``microscopic models'', where the TSLI or other internal variables are tracked for each individual in the population. Microscopic models can be more directly related to the mechanisms underlying the spreading of disease and to parameters measured in clinical studies. However, these models are too complex to understand the collective dynamics of an epidemic. Second, populations in reality (say, a local community relevant to the disease or an age group) are not infinitely large and thus exhibit fluctuations that reflect the infections of single individuals. Therefore, this raises the question of whether there exists an intermediate ``mesoscopic model'' that combines the simplicity of age-structured population models with the ability to capture fluctuations due to finite population size. Again, the parameters of such a mesoscopic model should be linked to the microscopic parameters. Answering these questions requires a multi-scale modeling framework for epidemic diseases.

We found our inspiration in the neuroscience field, where the age-structured population dynamics models have already been used in the form of the refractory density method \cite{ChizhovGraham2007,ChiGra08,DumHen24,DumTar24,gerstner_spiking_2002,PakPer09,SchwalgerChizhov2019,Tarn20,Schwalger2017}. Indeed, in that case, the basic units at the microscopic scale are the neurons, whose probability to fire a spike is largely determined by the time since the last spike as well as the spike input from other neurons in a large network of neurons. Taking into account the time since the last spike is important because neurons exhibit a period of absolute and relative refractoriness after a spike, where the firing probability is strongly reduced. At the macroscopic scale, the evolution of a neuronal population is governed by a system of first-order partial differential equations called refractory-density equations (RDE). The RDE tracks the density of the times since the last spikes within a continuous phase state, thereby avoiding discrete states such as spiking, refractoriness, and resting. Importantly, the RDE is a bottom-up mean-field model that is directly derived from the single neuron equations at the microscopic scale. Simulations using this technique yield precise solutions for both equilibrium states and transient non-equilibrium regimes.

In the last two decades, there have been several advancements in the refractory-density method that offered the possibility for significant progress in the multi-scale modeling of cortical networks in the brain. First, the method has been adapted for the important case where spiking events are triggered by the threshold crossing of a neuronal membrane potential that is driven by biologically plausible Gaussian white \cite{ChizhovGraham2007} or colored noise \cite{ChiGra08,Sch21}, while the classical RDE is based on a phenomenological hazard-rate-based spike-generation mechanism (``escape noise'') \cite{gerstner_spiking_2002}. Second, the RDE has been generalized to finite-size populations \cite{Schmutz2023,Schwalger2017}, resulting in a stochastic RDE that accurately reproduces the finite-size fluctuations at the mesoscopic scale. This advancement offers a unique and consistent description of a neural circuit at three levels of granularity: micro-, meso- and macroscopic scales. Additionally, there has been a further advancement in the generalization of RDE to accommodate complex, multi-dimensional \cite{ChizhovGraham2007}, and even two-compartmental \cite{ChiGra21} neurons, where the neuronal state is given by multiple variables rather than a single membrane potential. These advancements collectively allow the application of RDE for constructing meso- and macroscopic models based on a wide range of microscopic models.


The main concept of this paper is to adopt the recent generalizations of the refractory-density method from neuroscience to epidemiology. In particular, we suggest a novel multi scale modeling framework that connects three different scales (micro, meso, macro). Moreover, the theoretical framework applies to two different noise mechanisms at the microscopic scale (Gaussian white noise and escape noise). We show their distinct effects on the form of the age-structure epidemic population model at the meso- and macroscopic scales. To this end, we consider an infectious disease with a relatively short infection period, followed by temporary immunity and the possibility of multiple subsequent infections (\textit{e.g.}, flu or SARS-CoV-2). Our model can be naturally structured within a classical SEIRS (Susceptible -- Exposed -- Infected -- Recovered -- Susceptible) epidemic framework, where we continuously track the interplay between viral load and the immune system within each individual.

Unlike the classical ordinary differential equation (ODE)-based compartmental models of epidemics (see, \textit{e.g.}, \cite{li2002global}), we choose a continuous phase space, focusing on the time since infection and the corresponding evolution of the interaction between the virus and the immune system. In this space, the SEIRS stages can be defined either strictly or loosely; see section \ref{sec:individual} for a detailed explanation of our definition.
In reality, an individual does not experience strict boundaries between SEIRS stages, and our approach allows us to avoid such additional assumptions. Instead, we propose that two characteristics—- the time since the previous infection and the interplay between viral load and immune response -— sufficiently describe an individual's state within a population. We believe that avoiding strictly defined compartments and introducing a continuous space of states is a key advantage of our model.

Another advantage of our approach is that the spread of infectious disease at the mesoscopic and macroscopic population level is derived from the equations that stand for the dynamics of infection, disease development, and recovery of a single individual. A similar approach governs the derivation of macroscopic equations starting from their microscopic counterparts in kinetic theory, see \textit{e.g.} \cite{della2022sir,loy2021boltzmann}. Our construction differs however significantly, as we explain in sections \ref{sec:individual} and \ref{sec:population}. In particular, the present paper is, to the best of our knowledge, novel in that it combines such a multi-scale (microscopic, mesoscopic, macroscopic) model with an age (in the sense of ``infection age'') structured component, exploiting tools and analytical techniques coming from the so-called Refractory Density approach.

The paper is structured as follows: in Section \ref{sec:individual}, we present the microscopic model describing the evolution of the disease within a single individual. In Section \ref{sec:population} we introduce the equations governing the macro-scale population, while the mesoscopic ones are given in Section \ref{sec:meso}. We illustrate our work in Section \ref{sec:simulations} with numerical simulations and compare them to real data. We finally discuss our work in Section \ref{sec:discussion}.

\begin{figure}[p]
\centerline{\includegraphics[scale=0.6]{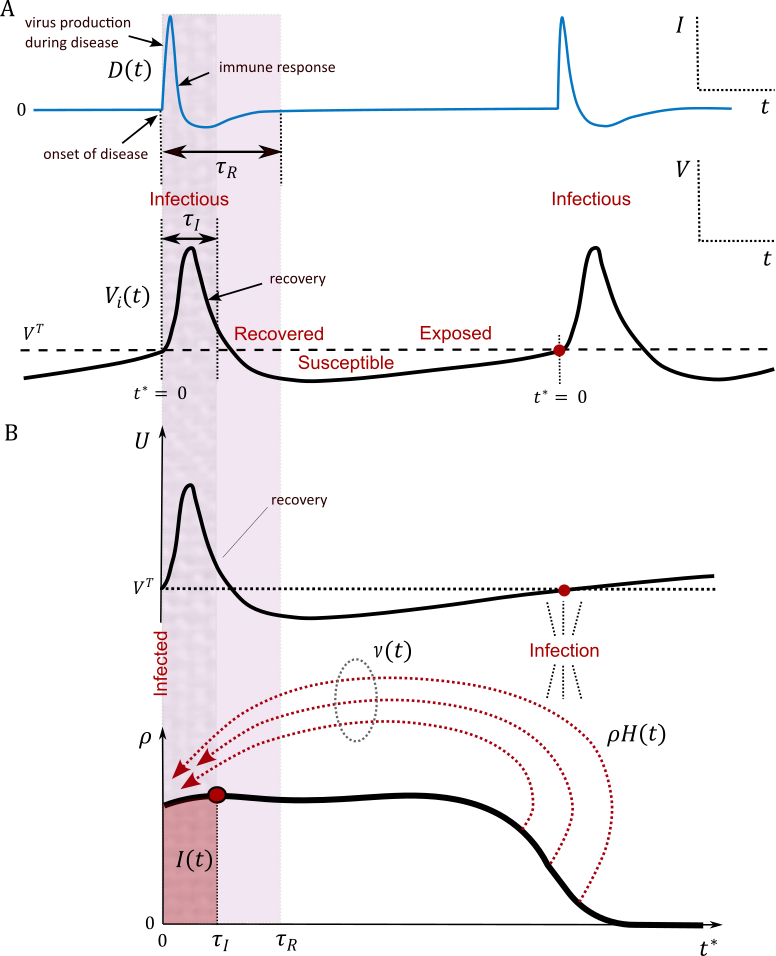}}
\caption{Schematic of the micro- and macroscopic models described in sections \ref{sec:individual} and \ref{sec:population}. {\bf A}, Microscopic model. The state variable $V_i$ of an individual $i$ evolves in time $t$ (bottom trace); its crossings of the threshold $\VT$ determine the onsets of disease, the infection time moments, when the drive $D(t)$ appear, which describes virus production and immune response (top). The time since the last infection $\ts$ is zeroed at the onset of the disease. {\bf B}, Macroscopic model. At time $t$, individuals with different $\ts$ are distributed with the density $\rho$ in the $\ts$-space (bottom). With time, the density transports rightwards (as $\ts$ increases with $t$) and undergoes a return flux to the infection state $\ts=0$ according to the hazard function $H$, so the dashed arrows represent subsequent infections. The hazard depends on 
the mean state variable $U$, which is similar to $V_i$ in {\bf A} but is extrapolated over the threshold.}
\label{fig:sketch}
\end{figure}

\section{Microscopic model of single individuals}\label{sec:individual}

We consider a population of $N\gg 1$ individuals, among which an infectious disease is spreading. We do not consider birth and death processes, and consequently, we assume the population to be constant (although this assumption is not critical for our framework).
We assume that the risk of infection for an individual depends on two state variables $V_i$ and $t^*_i$ (see Table \ref{Model Chart} for a summary of the variables and parameters of the model).
The variable $V_i$ represents the interplay of the viral load and the immune response within individual $i$. We call this variable \emph{viral state}. The dynamics of $V_i(t)$ are governed by the interaction of multiple processes (Fig.\ref{fig:sketch}A, bottom). 
First, due to the immune system, the number of viral particles relaxes with time constant $\tau$. 
Second, it grows proportionally to the fraction of infected individual $I(t)$, with a coefficient of proportionality $k$, which is assumed to be proportional to the basic reproduction number $R_0$ of the corresponding disease. This modeling assumption is qualitatively close to the underlying homogeneous mixing assumption of classical ODE models, whereby any individual is equally likely to meet with any other individual in the population. 
Third, it mimics the time course of the disease composed of a rapid increase of the viral load at disease onset and a subsequent decrease caused by the immune response. This time course is driven by the function $D(t_i^*)$, started at each onset of the disease (Fig.\ref{fig:sketch}A, top). 
Note that, since $V_i$ represents the interplay between the virus and the immune system, it can become negative at the negative phase of $D(t_i^*)$.

\begin{table}[t]
\centering
\begin{tabular}{|c|c|c|}
\hline
Variable & Notation & Units
\\ \hline
$t$ & Time & days \\
$\ts$ & Time since last infection, or ``infection age'' & days \\
$V_i$ & Interplay of viral load and immune response for individual $i$ & v-units \\
$U$ & Mean interplay of viral load and immune response & v-units \\
$\rho$ & Density of individuals distributed in $\ts$-space & people/day\\
$\nu$ & Population rate of new infections & people/day\\
$I$ & Fraction of infectious individuals in the population & nondim.\\
$\tau$ & Time scale of passive immune response & days    \\ 
$\tau_r$ & Time scale of disease progress and cure & days \\
$\tau_I$ & Infectious period & days    \\ 
$\tau_R$ & Time from infection to loss of immunity (``Refractory period'')& days \\
$D$ & Time course of the interplay for disease and treatment &  v-units/day   \\ 
$N$ & Total number of people in the population & people \\
$N_c$ & Cumulative cases in the population & people \\
$R_0$ & Basic Reproduction Number & nondim. \\
$k$ & Average individual rate of viral load & v-units/day\\
$H$ & Hazard; $H \Delta t$ is the probability of infection in the interval $(t,t+\Delta t)$  & people/day \\
$\VT$ & Threshold of infection & v-units \\
$\sigma$ & Noise amplitude & v-units \\
\hline
\end{tabular}
\caption{Notation for the micro- and macroscopic models described in sections \ref{sec:individual} and \ref{sec:population}.}
\label{Model Chart}
\end{table}

The state variable $t_i^*$ measures the time relative to the last infection time (onset of the disease) of individual $i$. These onsets define an ordered sequence $t_{1,i}<t_{2,i}<t_{3,i}<\dotsb$ of infection times for individual $i$. The continuous states of the epidemic in the $\ts$-space may be roughly interpreted in terms of discrete SEIRS stages (Fig.\ref{fig:sketch}A). To this end, we may partition the population with respect to an ongoing epidemic in the following compartments: $S$, individuals who are Susceptible to the disease; $E$, individuals who have been Exposed to the disease but are not yet infectious; $I$, Infected individuals who may spread the disease; and $R$, Recovered individuals who are temporarily immune from infection. According to these definitions, we can loosely define the stages in the $\ts$-axis. When $V_i(t)$ overcomes a threshold $\VT$ (in the case of Gaussian noise, see below) or an infection event is triggered by a probabilistic ``soft'' threshold (in the case of escape noise), the individual becomes infected and consequently belongs to compartment $I$. During the infectious phase, $V_i(t)$ rises above the threshold because of the viral production during disease and then decreases because of the immune response. The $I$ state lasts for a fixed duration $\tau_I>0$. After the time $\tau_I$ has elapsed since the onset of infection, the individual is effectively immune to the disease and consequently belongs to compartment $R$. The sojourn in this compartment lasts until the end of the recovery period $\tau_R>\tau_I$, measured from the onset of infection. After this time, when $V_i$ is low, the individual becomes Susceptible and then Exposed until the next infection when the individual becomes Infected and infectious again. Note that only the Infectious and Recovered states are strictly defined by the interval $0<\ts<\tau_I$ and $\tau_I\le \ts<\tau_R$, respectively. However, the duration of these states does not directly affect the evolution of the individual's state $V_i$, but instead determines the periods when the individual affects the population and is in an ``absolute refractory period'', respectively. 

Taken together, the dynamics for single individuals $i$, $i=1,\dotsc,N$, between two subsequent infection times $t_{k,i}$ and $t_{k+1,i}$ is modeled as 
\begin{align} \label{Eq1'}
    \frac{\text{d}V_i}{\text{d}t}&=-\frac{V_i}{\tau} 
    + k(t) I(t) + D(t_i^*) + \sigma(I(t))~ 
    \xi_i(t),\\
    \frac{dt_i^*}{dt}&=1,\\
\intertext{with the fraction of infected individuals}
    I(t)&=\frac{N_c(t)- N_c(t-\tau_I)}{N},\label{eq:I}       
\end{align}
where $N_c(t)\in\mathbb{N}$ is a global counting variable that keeps track of the cumulative number of cases in the population up to time $t$. Clearly, this counting variable is constant when no new infection occurs. At the infection times $t_{k,i}$, however, the overall dynamics is supplemented with an additional jump condition (reset): 
\begin{equation} \label{Eq2'}
    V_i(t_{k,i}^+)=V^T,\quad t_i^*(t_{k,i}^+)= 0,\quad N_{c}(t_{k,i}^+)= N_{c}(t_{k,i}^-)+1,
\end{equation}
where $t_{k,i}^-$ and $t_{k,i}^+$ denote the left- and right-sided limits, respectively. In Eq.~\eqref{eq:I}, we indicate with $\tau_I$ the \emph{average} time spent in the infectious state or \emph{average} duration of the infectious window. We note that Eq.~\eqref{eq:I} can be easily generalized to permit a more fine-grained, graded account of infectiousness depending on infection age $t^*$ (see Discussion, Eq.~\eqref{eq:I-generalized}). Furthermore,
\begin{align} \label{Eq21'}
    D(t^*) &= 4a \exp\left(-\frac{t^*}{\tau_r}\right) - a \exp\left(-\frac{t^*}{4 \tau_r}\right)  
\end{align}
is a response function that drives the time course of $V_i$ during disease.
The parameter $a$ is the severity of the disease, and $\tau_r$ is the time of disease progress and cure.
The last term in Eq.~\eqref{Eq2'} represents an independent Gaussian white noise with mean $\mathbb{E}[\xi(t)]=0$, auto-covariance function $\mathbb{E}[\xi(t)\xi(t')]=\delta(t-t')$ and strength $\sigma(I)$ ($\delta(\cdot)$ denotes the Dirac delta function). As explained below, this term is only present in our model variant with Gaussian noise and is absent in the variant with escape noise where we set $\sigma(I)\equiv 0$.

As initial conditions of the model, we assume that before time $t=0$ there were no cases, $N_c(t)=0$ for $t<0$ and that at time $t=0$ a fraction $I_0$ of the population gets infected, \textit{i.e.} $N_c(0)=I_0 N$. Thus the state variables for the $I_0 N$ initially infected individuals are $V_i(0)=V^T$ and $t_i^*(0)=0$, while the state variables of all other (non-infected) individuals (the remaining $(1-I_0)N$ individuals in the population) is initially $V_i(0)=0$ and $t_i^*(0)=+\infty$. 

Importantly, it remains to define how the infection events are triggered. We consider two different variants to define these events, and hence the infection times $t_{k,i}$, corresponding to two different ways of including stochasticity in the model. We refer to these variants as the \emph{white}- and \emph{escape}-noise cases, respectively.

\paragraph{Model with escape noise}

To introduce stochasticity that phenomenologically captures the influences of various sources of noise, we model the infection events of an individual $i$ by a probabilistic risk of becoming infected depending on the individual's current state variables $V_i(t)$ and $t^*_i(t)$. Mathematically, we keep the dynamics of $V_i$ between infection events deterministic (\textit{i.e.} setting $\sigma=0$) and generate infection events stochastically by a state-dependent hazard rate
\begin{equation}    \lambda_i(t)=H(V_i(t),t_i^*(t)).
\end{equation}
The hazard rate means that an infection of individual $i$ occurs in a small time step of length $\Delta t$ with conditional probability $H_i(t)\Delta t$ given the individual's current state $(V_i(t),\ts_i(t))$. More precisely, we have the following conditional probability:
\begin{equation} \label{Eq2''}
\text{Pr}\left\{\text{individual $i$ gets infected in }(t,t+\Delta t)\;|\;V_i(t),t_i^*(t^-)\right\}=\lambda_i(t)\Delta t+o(\Delta t)
\end{equation}
as $\Delta t\to 0$.
For a concrete choice of the hazard rate in simulations, we use a rectified power function with absolute refractory period $\tau_R$: 
\begin{equation}  \label{Eq_escape}
H(V,t^*)=c\max(0,V/\VT)^m\mathbbm{1}_{t^*\ge\tau_R},
\end{equation}
where $m>0$. We illustrate the effect of varying $m$ in Figure \ref{fig:U_nu5}.

\paragraph{Model with white Gaussian noise} 
An alternative way to model noise is to consider 
an additive white-noise drive in the dynamics of $V_i$. This ``diffusive noise'' is modeled by the term $\sigma(I(t))~ 
\xi_i(t)$ in Eq. \eqref{Eq1'}. The stochastic term may reflect both the fluctuations of the number of viral particles of a given type around a particular individual and the fluctuations of activity of the immune system of this individual. To prevent that in the absence of a viral infection in the population an infection occurs spontaneously by noise, we enforce the noise strength to be zero when $I(t)$ is zero by choosing the specific function $\sigma(I)=\hat{\sigma}\mathbbm{1}_{I>0}$. In the white-noise case, infection events are triggered when the viral state $V_i(t)$ crosses a certain threshold $V^T$, provided that at least an absolute refractory time $\tau_R$ has passed since the last infection. Hence, the infection times $t_{k,i}$ are defined by the condition
\begin{equation}
    V(t_{k,i}^-)\ge V^T,\quad t^*(t_{k,i}^-)\ge \tau_R.\label{eq:thresh}
\end{equation}
Note that if this condition is fulfilled at some time $t_{k,i}^-$, the reset condition Eq.~\eqref{Eq2'} ensures that the above threshold condition no longer holds at $t_{k,i}^+$, \textit{i.e.} immediately after this time, as it should be.

The above equations conclude the definition of the microscopic model. Given the following application of the refractory-density (RD) method, we note that the case of white Gaussian noise can be approximately mapped to the case of escape noise \cite{ChizhovGraham2007,Sch21}. In this case, the noise term in Eq. \eqref{Eq1'} is omitted and the condition Eq.~\eqref{eq:thresh} is substituted by \eqref{Eq2''}, where $H_i(t)$ is now given by the following hazard function (\ref{Eq8}). The hazard rate $\lambda(t,\ts)$ depends on the actual state $V_i$, its rate of change $\text{d}V_i/\text{d}t$, the threshold $\VT$, and the noise  amplitude  $\sigma$, i.e., $\lambda(t,\ts)=H(V_i(t),\dot V_i(t),\ts_i(t); \sigma(I(t)))$. In \cite{ChizhovGraham2007}, this function was found as an approximate solution of a first-time passage problem based on the Kolmogorov-Fokker-Planck equation:
 \begin{align}  \label{Eq8}
H(U,\dot U,\ts,\sigma)&=\lbrace A(T)+B(T,\dot U,\sigma), ~~\hbox{if}~~ \ts>\tau_R, ~~\sigma>0; ~~~0, ~\hbox{otherwise} \rbrace,  
\\
A(T)&= \frac{1}{\tau} ~\exp (0.0061-1.12 T - 0.257 T^2 - 0.072 T^3 - 0.0117 T^4),
\nonumber\\
B(T,\dot U,\sigma)&= \frac{2}{\sigma \sqrt{\pi}} \bigg\lbrack \dot U \bigg\rbrack_{+} \frac{\exp(-T^2)}{1+\hbox{erf}(T)}, 
~~~~T=\frac{\VT-U}{\sigma}.\nonumber
\end{align}
Here and in the following, $[x]_+=\max(0,x)$ denotes the rectified linear function. We refer to \cite{Sch21} for a slightly more elaborate approximation of diffusive noise by escape noise.
\section{Macroscopic model of a population}\label{sec:population}

For a large number of individuals $N$, the dynamics of the microscopic model can be either evaluated with a Monte-Carlo simulation or obtained much more efficiently by integrating corresponding macroscopic population equations.
To this end, we apply the refractory-density (RD) approach \cite{SchwalgerChizhov2019} to the epidemiological model described at an individual level in section \ref{sec:individual}. 
According to this approach, the state variables of a single ``particle''/individual are parameterized by the ``age'' $\ts$ (with the total derivative in time substituted by the sum of partial derivatives, i.e. $\text{d}/\text{d}t=\partial /\partial t + \partial/\partial \ts$), and the population is characterized by the density of particles in the one-dimensional, semi-infinite $\ts$-space, i.e.  $\rho(t,\ts)$ (Fig.\ref{fig:sketch}B). In our case, in line with other age-dependent models in mathematical epidemiology, the ``age'' is the time elapsed since infection. This means that at any time $t$, the number of individuals in the population who became infected between $\ts_1$ and $\ts_2$ time units ago are represented by
$$
\int_{\ts_1}^{\ts_2} \rho(t,\ts)\text{d}\ts.
$$
At the macroscopic level, we define the rate of new infections
\begin{equation} \label{Eq3''}
    \nu(t)=\lim_{\Delta t \to 0}\lim_{N \to {} \infty} \frac{\Nc(t+\Delta t)-\Nc(t)}{N \Delta t}.
\end{equation}
We also introduce a function $U(t,\ts)$ as follows: for the model with escape noise, $U(t,\ts)$ is the unique function that maps the infection age $\ts_i(t)$ to the state variable $V_i(t)$, \textit{i.e.} $V_i(t)=U(t,\ts_i(t))$. For the model with Gaussian white noise, there is no such deterministic map. However, we can first map the model to the approximate escape-noise model, Eq.~\eqref{Eq8}, for which we can define again the unique function $U(t,\ts)$. Intuitively, this function can be interpreted 
as the variable $V_i(t)$ averaged across realizations of the Gaussian-white noise for individuals $i$ that had the time of last infection equal to $t^*$, \textit{i.e.}, the same $\ts$. The equation for $U$ then follows from the corresponding averaging of Eq. \eqref{Eq1'}.
The bottom traces in Fig.\ref{fig:sketch}B illustrate distributions of $U$ and $\rho$ in $\ts$-space.

The RD model corresponding to Eqs. \eqref{Eq1'}-\eqref{Eq2'} consists of two transport equations for $\rho(t,\ts)$ and $U(t,\ts)$, and two integrals for $\nu(t)$ and $I(t)$:
\begin{subequations}
\label{eq:macro}
\begin{align}  \label{Eq45'}
\derc {\rho}t + \derc {\rho}{\ts}&=-\rho ~H(U(t,\ts),t^*), \\
\derc Ut + \derc U{\ts}&=-\frac{U}{\tau} 
+ k(t)I(t) + D(\ts), \label{Eq46'}\\
\nu(t)&= \int_0^{\infty} \rho(t,\ts) ~H(U(t,\ts),\ts) \,d\ts + I_0 \delta(t), & \text{rate of new cases} \label{Eq48'}\\
I(t)&=
\int_{t-\tau_I}^{t} \nu(s) \,ds, & \text{fraction of infected people} \label{Eq47'}\\
\end{align}
\end{subequations}
The boundary condition for $\rho$, resulting from the conservation of people in the population, is 
\begin{equation}
   \rho(t,0)=\nu(t). \label{Eq49'} 
\end{equation}
The boundary condition for $U$ corresponding to the reset in the microscopic model, Eq.~\eqref{Eq2'}, is $U(t,0)=V_T$.
The initial condition for the density $\rho$ corresponding to an initial fraction of infected people $I_0$ at $t=0$ and to a situation, where individuals never encountered this viral infection before time $t=0$ (formally, $\ts_i(0^-)=\infty$) is
\[\rho(0,\ts)=\delta(\ts) I_0 + \delta(\ts-\infty) (1-I_0).\]
Furthermore, the corresponding initial condition for $U$ can be taken as the steady-state solution of Eq.\eqref{Eq46'} (with $\partial U/\partial t=0$ and $I(t)=0$ for $t<0$):
\[U(0,\ts)=\VT \exp{(-\ts/\tau)} +\int_0^{t^*}e^{-s/\tau}D(\ts-s)\,ds.\] 
In the density equation \eqref{Eq45'}, the density diminishes because of the sink term $-\rho H$ but the total mass $\int_0^\infty\rho(t,\ts)\,d\ts=1$ is conserved at all times because of the boundary condition, Eq. \eqref{Eq49'}. This boundary condition acts as a source term at $\ts=0$ given by the rate of new cases $\nu(t)$ that exactly balances the integrated sink term, Eq.~\eqref{Eq48'}. We note that this conservation law reflects the fact that we neglect mortality, and focus on the transmission mechanism.  

Furthermore, in the density equation \eqref{Eq45'}, the risk of illness is evaluated by the hazard function $H$. In Eq.~\eqref{eq:macro}, we used the hazard function for the case of escape noise.
For the case of white noise, the hazard function $H(U(t,\ts),\ts)$ should be replaced by $H(U(t,\ts),\dot U(t,\ts),\ts,\sigma(I(t)))$, where $H$ is given by Eq. \eqref{Eq8} and $\dot U(t,\ts)=(\partial_t+\partial_{t^*}) U(t,t^*)$ is given by the right-hand-side of Eq.~\eqref{Eq46'}. 

Once again,  note that Eq.~\eqref{Eq47'} permits a simple generalization to graded infectiousness depending on the course of the infection (see Discussion, Eq.~\eqref{eq:I-generalized}). 

\section{Mesoscopic model of a finite-size population} \label{sec:meso}


At the intermediate mesoscopic scale, the size of (sub-)populations is finite. The finite size causes fluctuations in the fraction of infectious people $I(t)$, which may yield significant finite-size effects through the interaction of finite-size noise and nonlinear population dynamics. In the case of a finite size population consisting of $N \in \mathbb{N}$ individuals, $N\gg 1$, we apply the corresponding theory from \cite{Schmutz2023,Schwalger2017}, which yields a stochastic generalization of the macroscopic dynamics given in the previous section. To this end, we introduce the pseudo-density $\rho(t,t^*)$ in terms of the survivor function $S(t,t^*)$ as
\begin{subequations}
\begin{equation}
    \rho(t,t^*)=S(t,t^*)~\nu_N(t-t^*). \label{EqMeso1}
\end{equation}
Here, $S$ and $\nu_N$ are given as the solution of the following system of stochastic partial differential equations, generalizing Eq.~\eqref{eq:macro}
\begin{align}  \label{Eq45''}
\derc {S}t + \derc {S}{\ts}&=-S ~H(U),\qquad S(t,0)=1, \\
\derc Ut + \derc U{\ts}&=-\frac{U}{\tau} + u(t) + k~I(t) + D(\ts),\qquad U(t,0)=V_T, \label{Eq46''}\\
I(t)&=\int_{t-\tau_I}^t \nu_N(s) \,ds,  \label{Eq47''}\\
\nu(t)&= \left[\int_0^{\infty} \rho(t,\ts) ~H(U(t,\ts)) \,d\ts+\bar H(t)\left(1-\int_0^\infty\rho(t,t^*)\,dt^*\right)\right]_+, \label{Eq48''}\\
\nu_N(t)&=\nu(t)+\sqrt{\frac{\nu(t)}{N}}\xi(t), \label{Eq49''}
\end{align}
with
\begin{equation} \label{EqMeso7}
    \bar H(t)=\frac{\int_0^\infty H(U(t,\ts))(1-S(t,t^*))\rho(t,t^*)\,dt^*}{\int_0^\infty (1-S(t,t^*))\rho(t,t^*)\,dt^*},
\end{equation}
\end{subequations}
where $\xi(t)$ is the white Gaussian noise of unity amplitude. We remark that, for all $t,t^*$, $S(t,t^*), \in [0,1]$; hence, the function $\bar H(t)$ is non-negative for all $t$. Furthermore, we note that $\nu_N(t)$ must be regarded as an abstract quantity that mathematically only makes sense as a distribution, \textit{i.e.} if it is integrated against some test function over time. Specifically, the empirical rate of new cases $\hat{\nu}_N$ measured over some finite time step $\Delta t$ (such that the expected number of new cases $N\nu(t)\Delta t\gg 1$) would be
\begin{equation}
    \hat{\nu}_N(t)=\left[\int_t^{t+\Delta t}\nu_N(s)\,ds\right]_+.
\end{equation}
While in practice it is extremely rare that the integral takes negative values for large $N$, we added the rectification $[\cdot]_+$ to enforce a non-negative rate. The same statement also holds true for Eq.~\eqref{Eq48''}.

Likewise, $\rho$ is not a normalized probability density \cite{Schmutz2023,Schwalger2017}. Therefore, it is no longer the strict interpretation of the empirical density of the times since the last infection but $\rho(t,t^*)dt^*$ must be regarded as an abstract measure. 
In comparison with the macroscopic model, this system of equations is stochastic. Its solution corresponds to a single realization of the noise $\xi(t)$. When  $N \xrightarrow{} \infty$, the solution converges to the solution of the macroscopic model.
The precise numerical method employed for these simulations is described in \cite{Schmutz2023} (see also the provided simulation code in Python). 


\section{Simulations}\label{sec:simulations}

\begin{figure}[h!]
\centering
\hbox{A ~~~~~~~~~~~~~~~~~~~~~~~~~~~~~~~~~~~~~~~~~~~~~~~~~~~~~~~~~~~~~~~~~~~~~~~~~~~~~~~~~~~~~B}
\centerline{\includegraphics[scale=0.14]{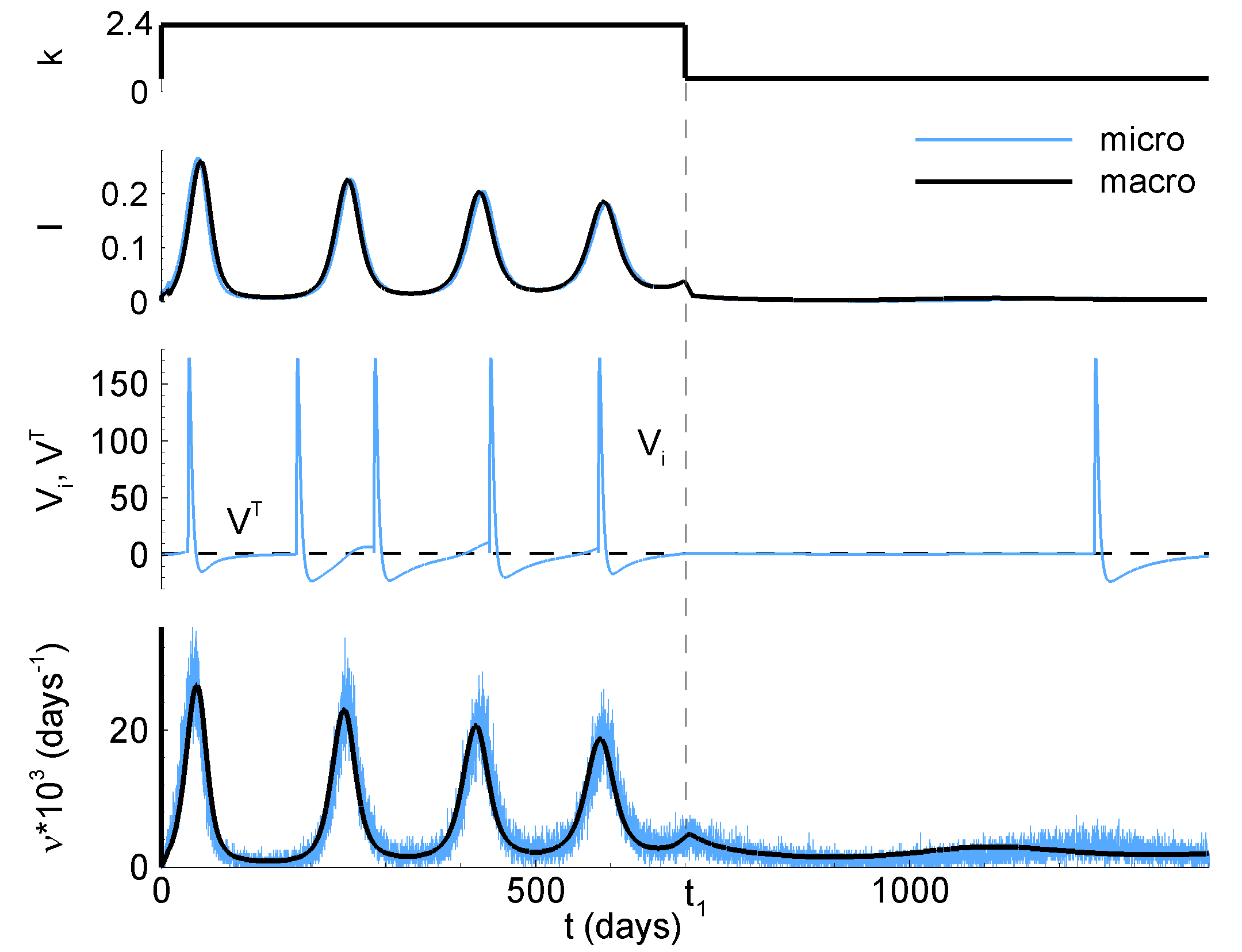}
            \includegraphics[scale=0.08]{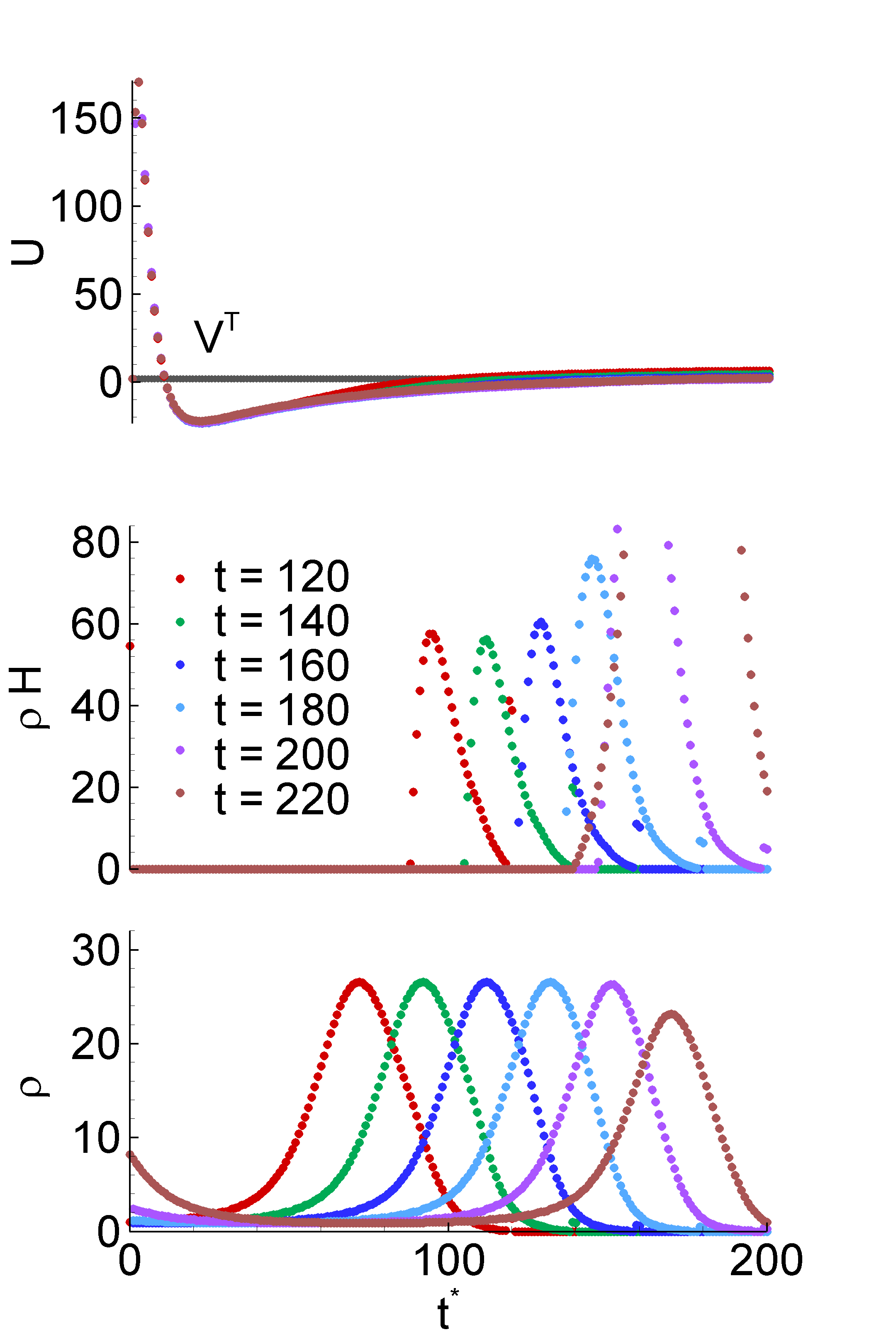}}
\caption{Escape noise in the form of \eqref{Eq_escape}. Simulation of the response to the rapid change of the rate of viral load $k$ from 0.6 to 2.4 and the short infecting pulse $I_0(t)$, at $t=100$. {\bf A}, top to bottom: $I_0(t)$, $k(t)$, the fraction of infected individuals $I(t)$, the state variable $V_i(t)$ of a representative individual, and the rate of new cases $\nu(t)$. 
For $I(t)$ and $\nu(t)$, the solutions obtained with both macro- (black lines) and microscopic equations for $N=20000$ (blue lines) are shown. {\bf B}, the distribution in the $\ts$-space of the mean state variable $U$, the source term $\rho H$, and the density $\rho$, in 4 time moments. 
Parameters: 
$m=1$, $c=0.015$~days$^{-1}$,
$\tau=50$, $\tau_I=10$, $\tau_r=1$, $\tau_R=20$, $a=100$, and $\VT=2$. }
\label{fig:U_nu5}
\end{figure}

\paragraph{Model with escape noise}
A first simulation of epidemic waves is shown in Fig. \ref{fig:U_nu5}.  
The onset of an epidemic is modeled 
by a short pulse of magnitude $I_0$ of the rate of new infections $\nu(t)$ at time $t=0$ reflecting the appearance of contaminating patients. These conditions result in oscillations of the rate of cases $\nu(t)$ and the fraction of potentially infected population $I(t)$. These oscillations tend to relax. The dynamics of a single individual state $V_i(t)$ (Fig. \ref{fig:U_nu5}A, next to bottom panel) follows the epidemic waves of $\nu(t)$ and $I(t)$. Each peak of $V_i(t)$ reflects the virus reproduction during the disease. The decreasing phase reflects the immune response, after which an individual becomes again susceptible and might get infected during the next contaminating flux $k(t) I(t)$. Each spike of the rate of cases $\nu(t)$ leads to a correspondingly shaped distribution of the density in the $\ts$-space.

The population waves move in the $\ts$-space from $\ts=0$ towards the region where $U$ is about to cross the threshold $V^T$ and form the peaks of $\rho H$, which is the distribution of falling sick individuals (Fig. \ref{fig:U_nu5}B). Since the integral over $\rho H$ determines the rate of cases $\nu(t)$, the increase of the $\rho H$ determines the next peak of $\nu(t)$ and $I(t)$, and so on. 

We assume that at $t=t_1=700$ the basic reproduction rate drops 4 fold, which might be due to any kind of containment measures. As a result, the epidemic wave generation stops through a significant decrease in the rate of cases $\nu(t)$ (Fig. \ref{fig:U_nu5}A).

\begin{figure}[h!]
\centerline{\includegraphics[scale=0.14]{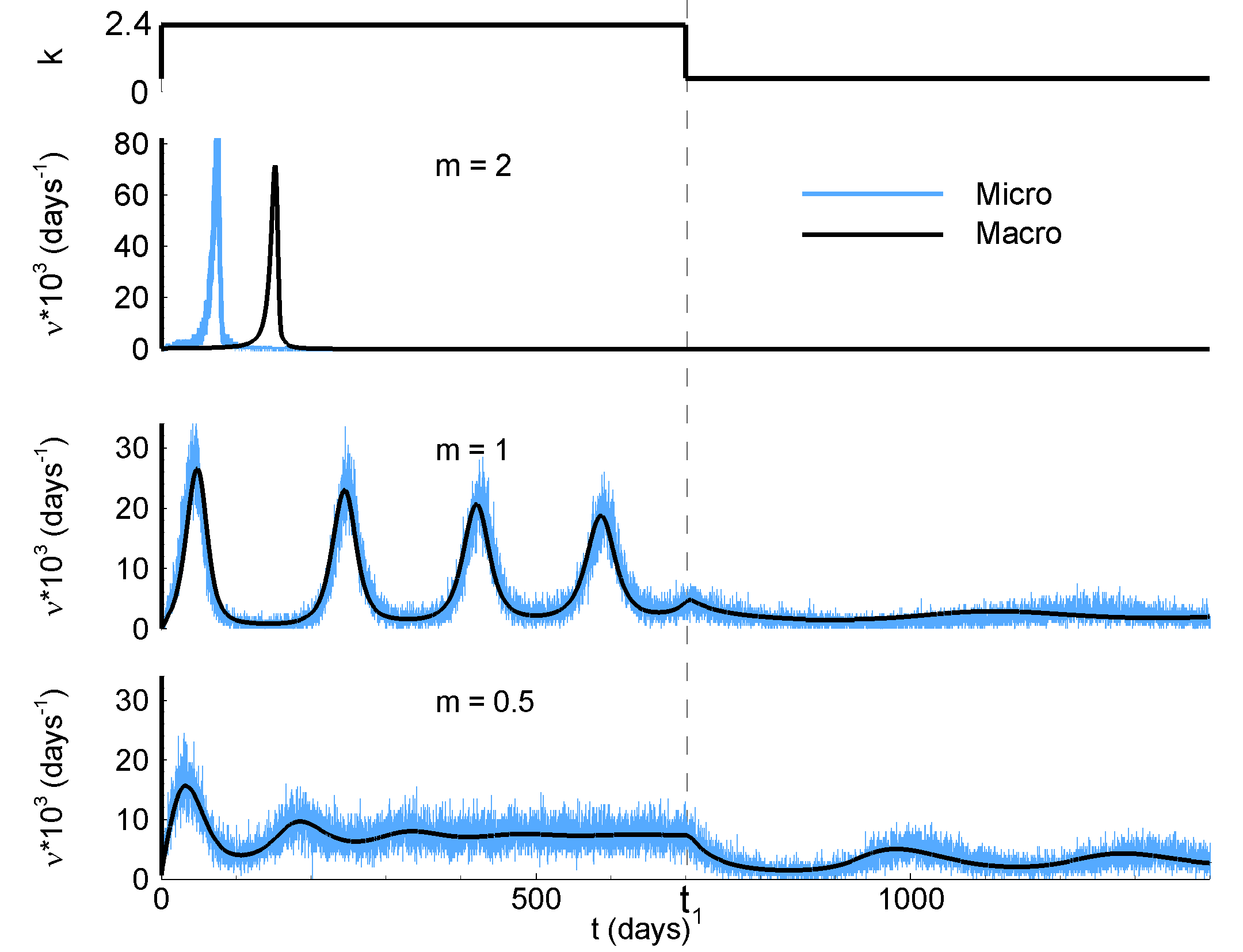}}
\caption{Escape noise in the form of \eqref{Eq_escape} with $m=$ 0.5, 1 and 2. The remaining parameter values are as in Fig. \ref{fig:U_nu5}.}
\label{fig:nu567}
\end{figure}

For the escape noise in the form of \eqref{Eq_escape}, the epidemic depends on the steepness  $m$ of the hazard function (Fig. \ref{fig:nu567}). With the increase of $m$ from 0.5 to 1, the epidemic oscillations increase. However, it is non-trivial that in the case of more steep $H$ with $m=2$ the epidemic shows only one peak. In this case, the rapid illness of the whole population results in a short splash of contamination $k(t) I(t)$ that occurs at the recovery phase of the population when no one is susceptible.

\begin{figure}[h!]
\hbox{A}
\centerline{\includegraphics[scale=0.14]{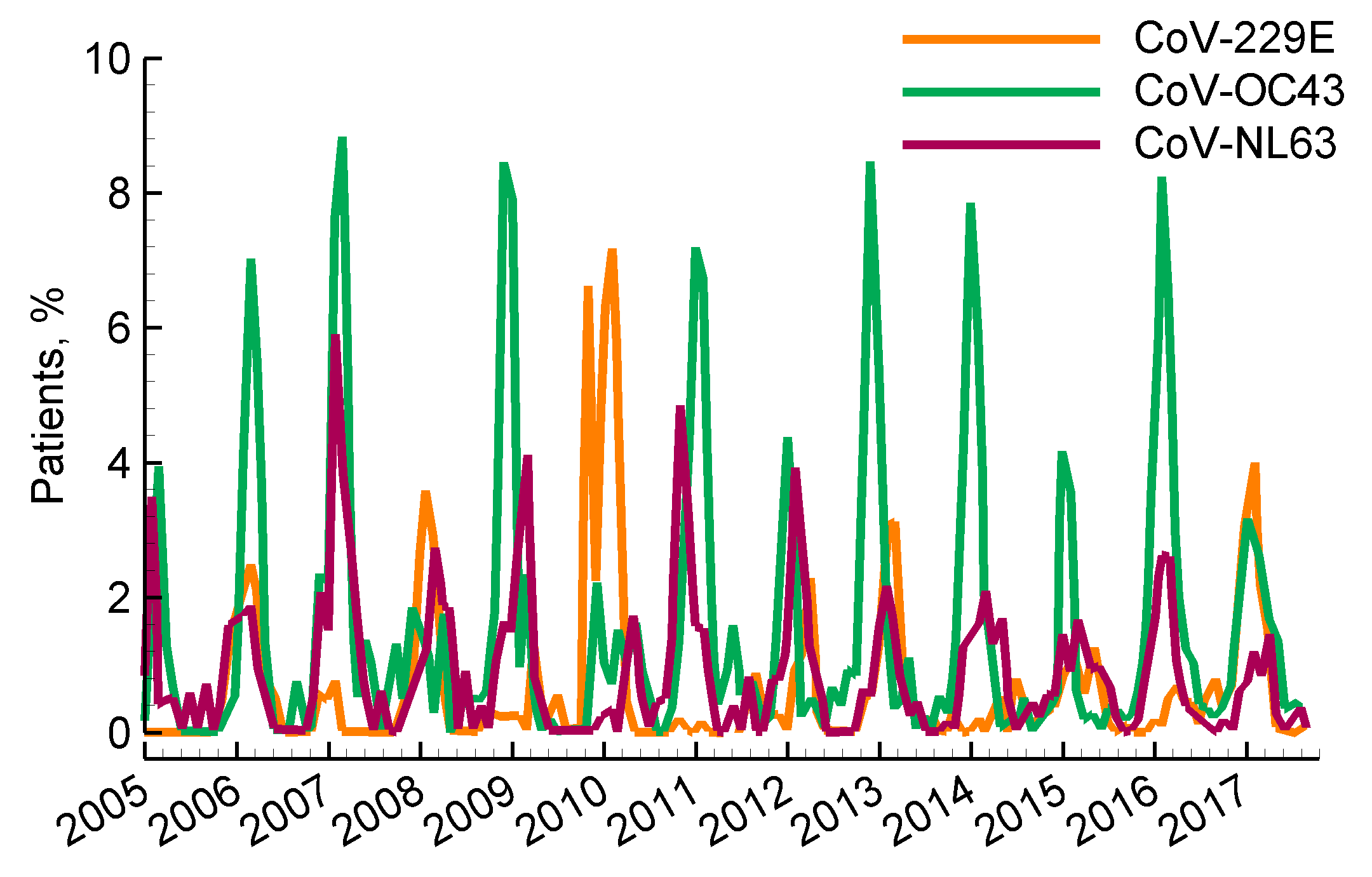}}
\hbox{B}
\centerline{\includegraphics[scale=0.14]{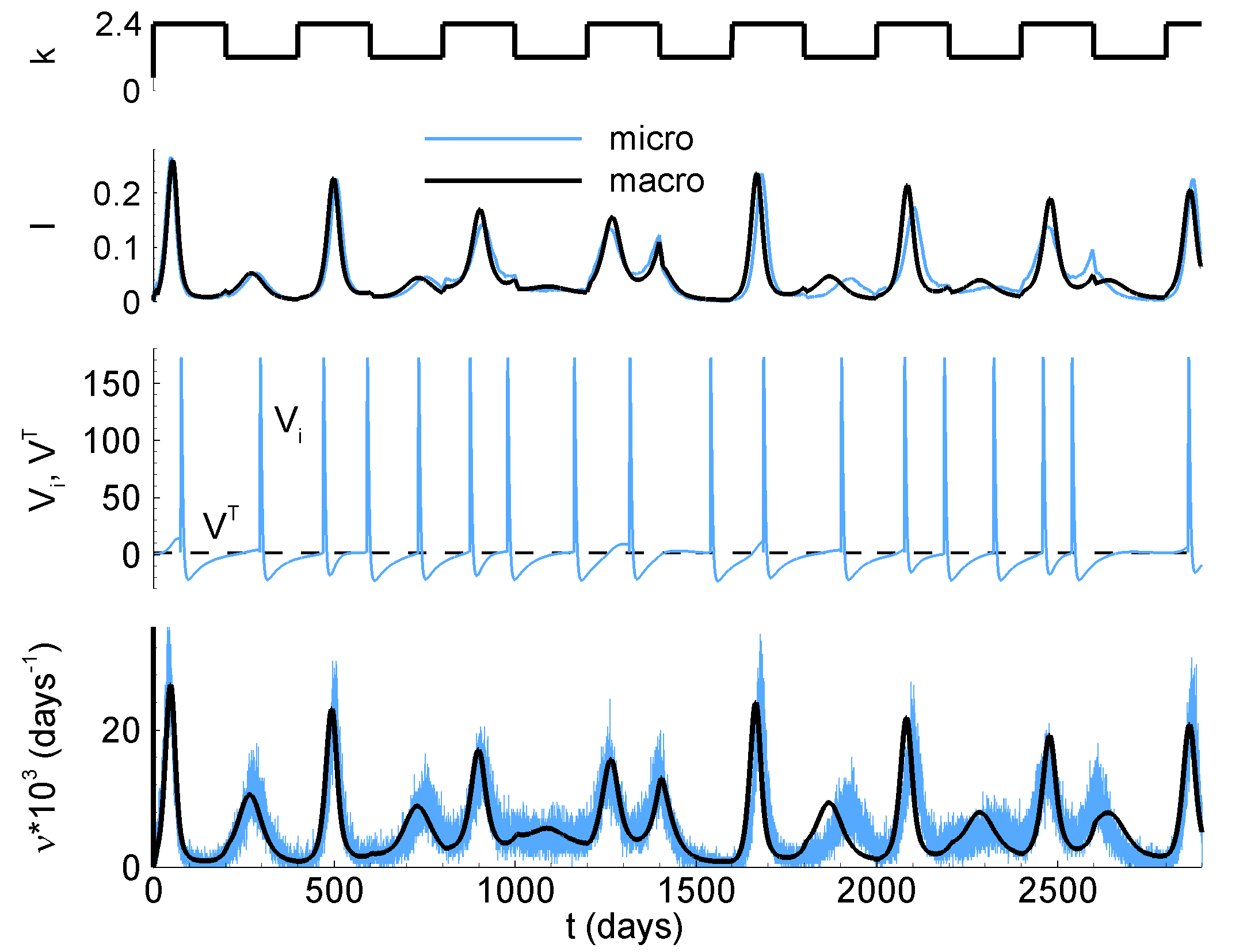}}
\caption{Epidemiology of Seasonal Coronaviruses. {\bf A}, Data from \cite{Nickbakhsh2020}; specifically, the monthly prevalence of seasonal coronaviruses (sCoVs) detected among
patients with respiratory illness virologically tested in NHS Greater Glasgow and Clyde, Scotland, United Kingdom.
{\bf B}, The case of oscillating $k$. Escape noise in the form of \eqref{Eq_escape}. The remaining parameter values are as in Fig. \ref{fig:U_nu5}.  
}
\label{fig:U_nu8}
\end{figure}

\paragraph{Seasonality}
One of the experimentally observed features of epidemics is its seasonal oscillations, as illustrated in Fig. \ref{fig:U_nu8} by data from \cite{Nickbakhsh2020}. In Fig. \ref{fig:U_nu8}, we illustrate the case of seasonal change of $k$ through a step function alternating between $k=2.4$ and $k=1.2$ every 180 days. In response to a meander-like $k$, we observe a complex pattern of waves with various amplitudes.  

\begin{figure}[h!]
\centering
\hbox{A ~~~~~~~~~~~~~~~~~~~~~~~~~~~~~~~~~~~~~~~~~~~~~~~~~~~~~~~~~~~~~~~~~~~~~~~~~~~~~~~~~~~~~B}
\centerline{\includegraphics[scale=0.14]{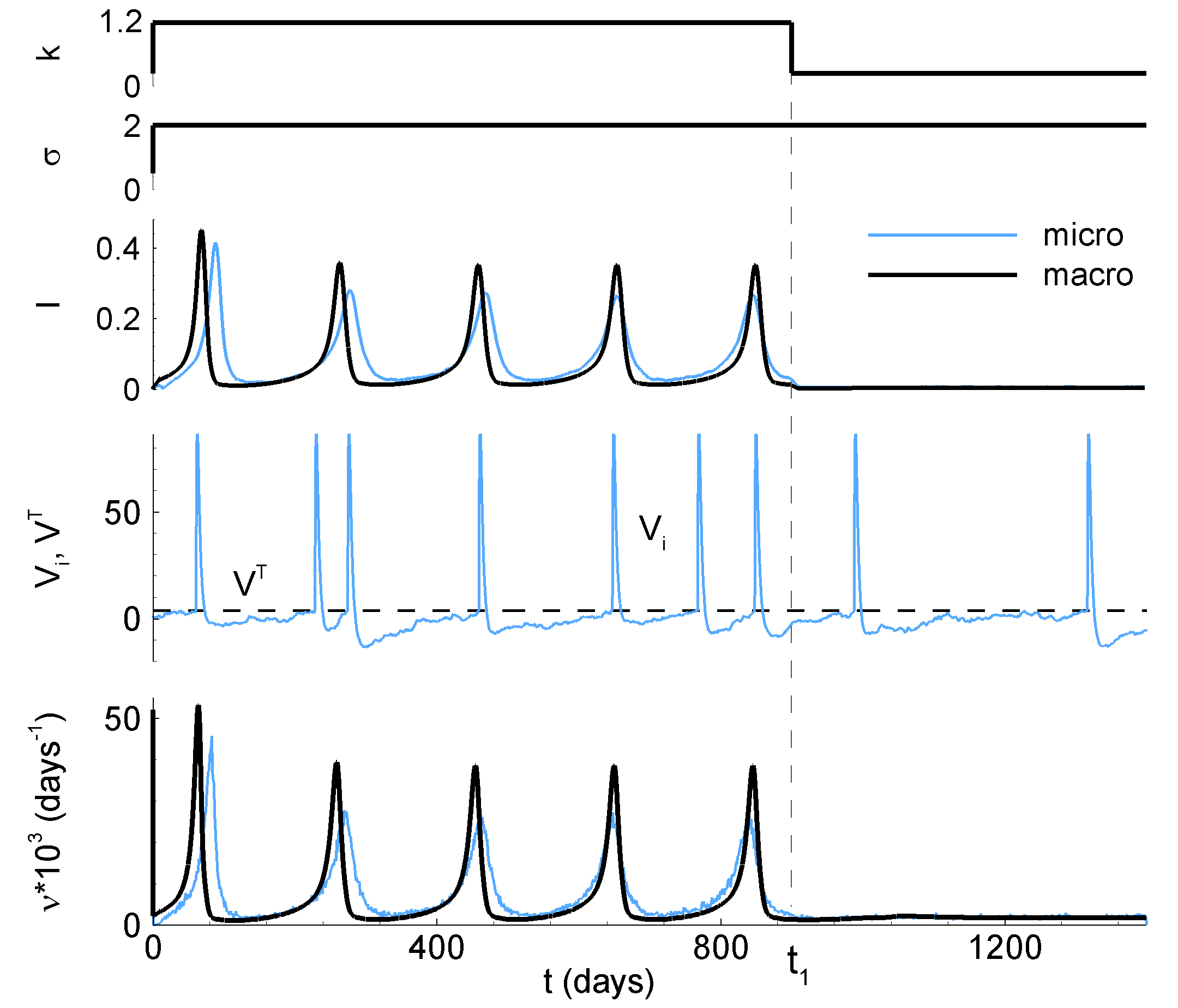}
            \includegraphics[scale=0.08]{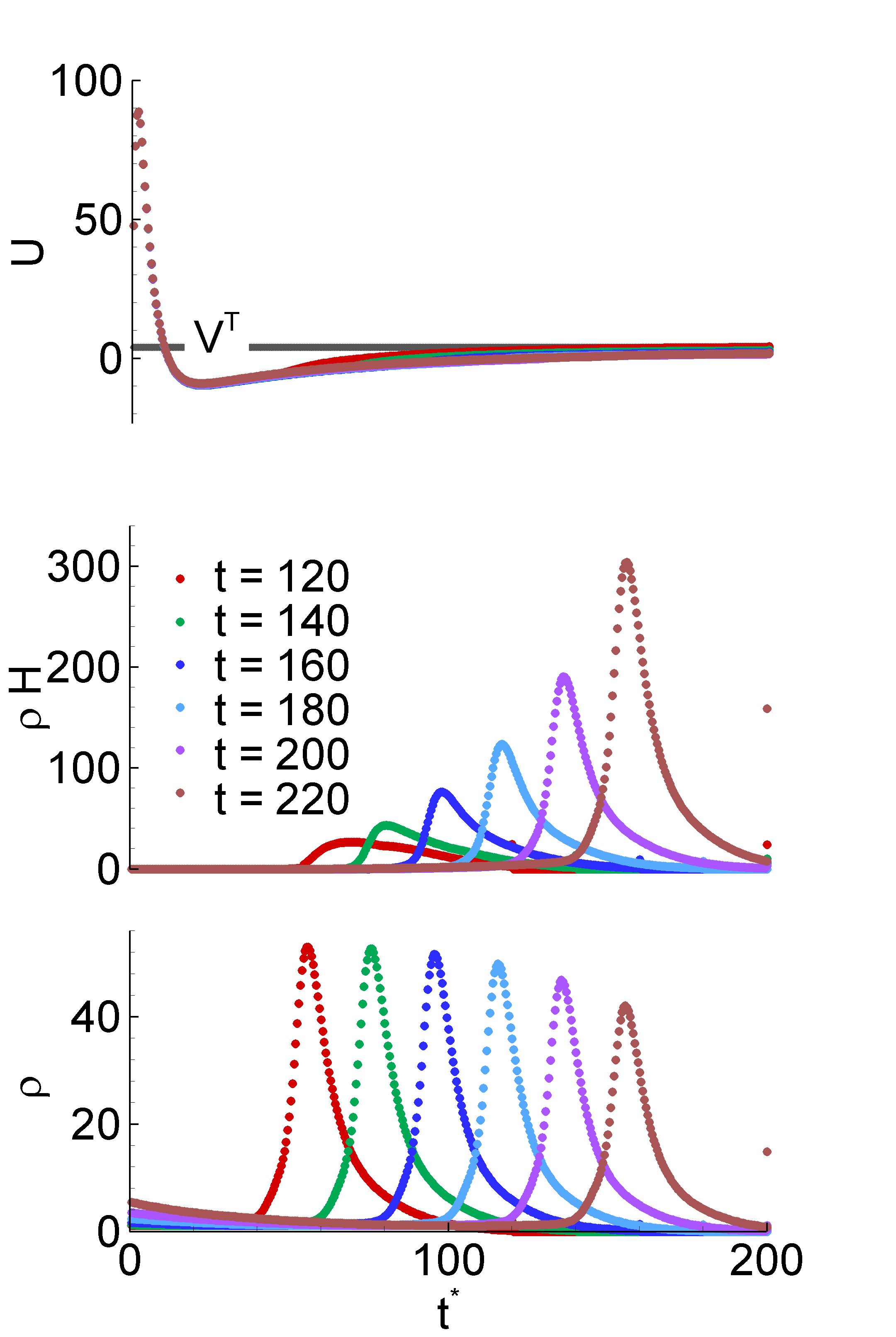}}
\hbox{C ~~~~~~~~~~~~~~~~~~~~~~~~~~~~~~~~~~~~~~~~~~~~~~~~~~~~~~~~~~~~~~~~~~D}
\centerline{\includegraphics[scale=0.105]{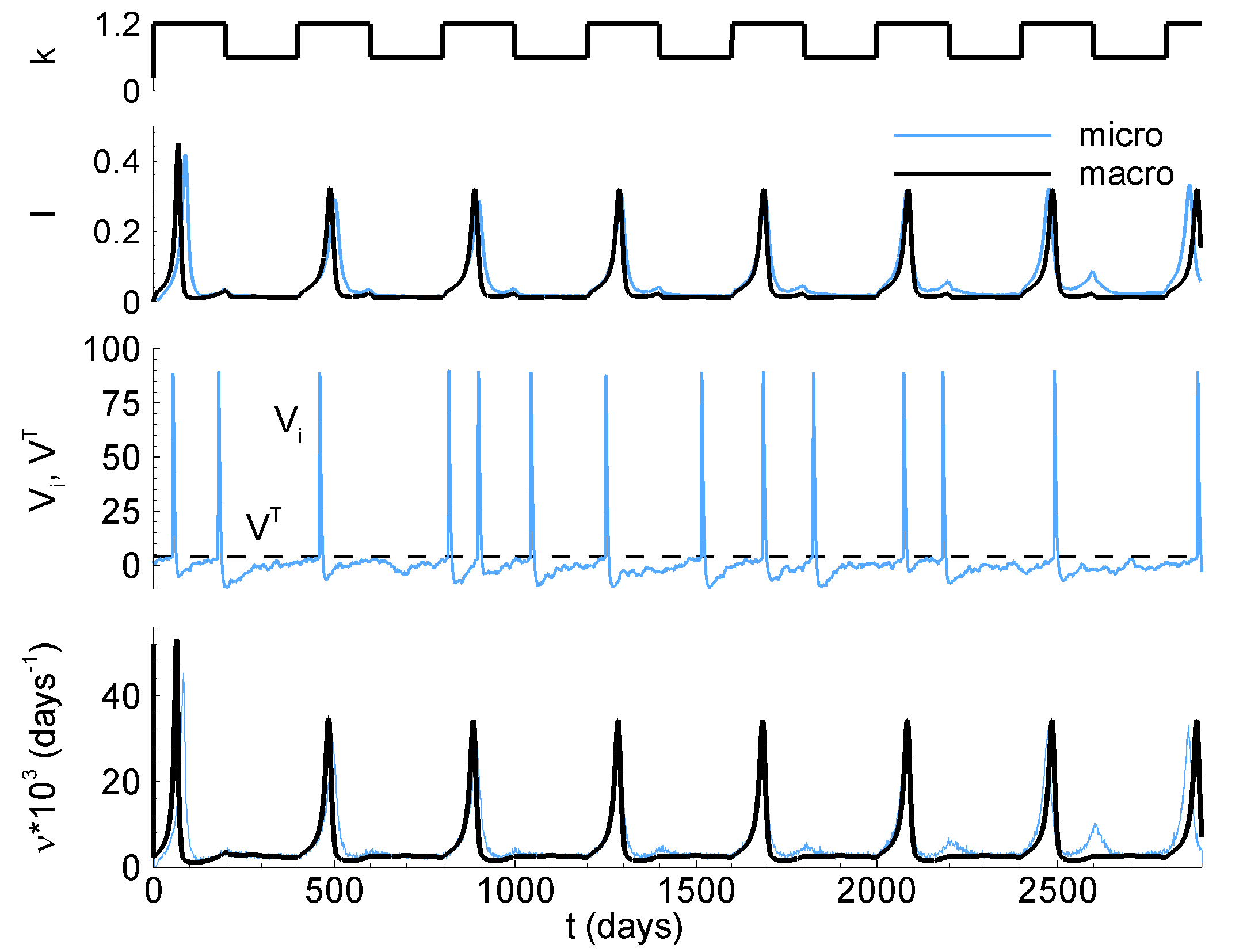}
            \includegraphics[scale=0.105]{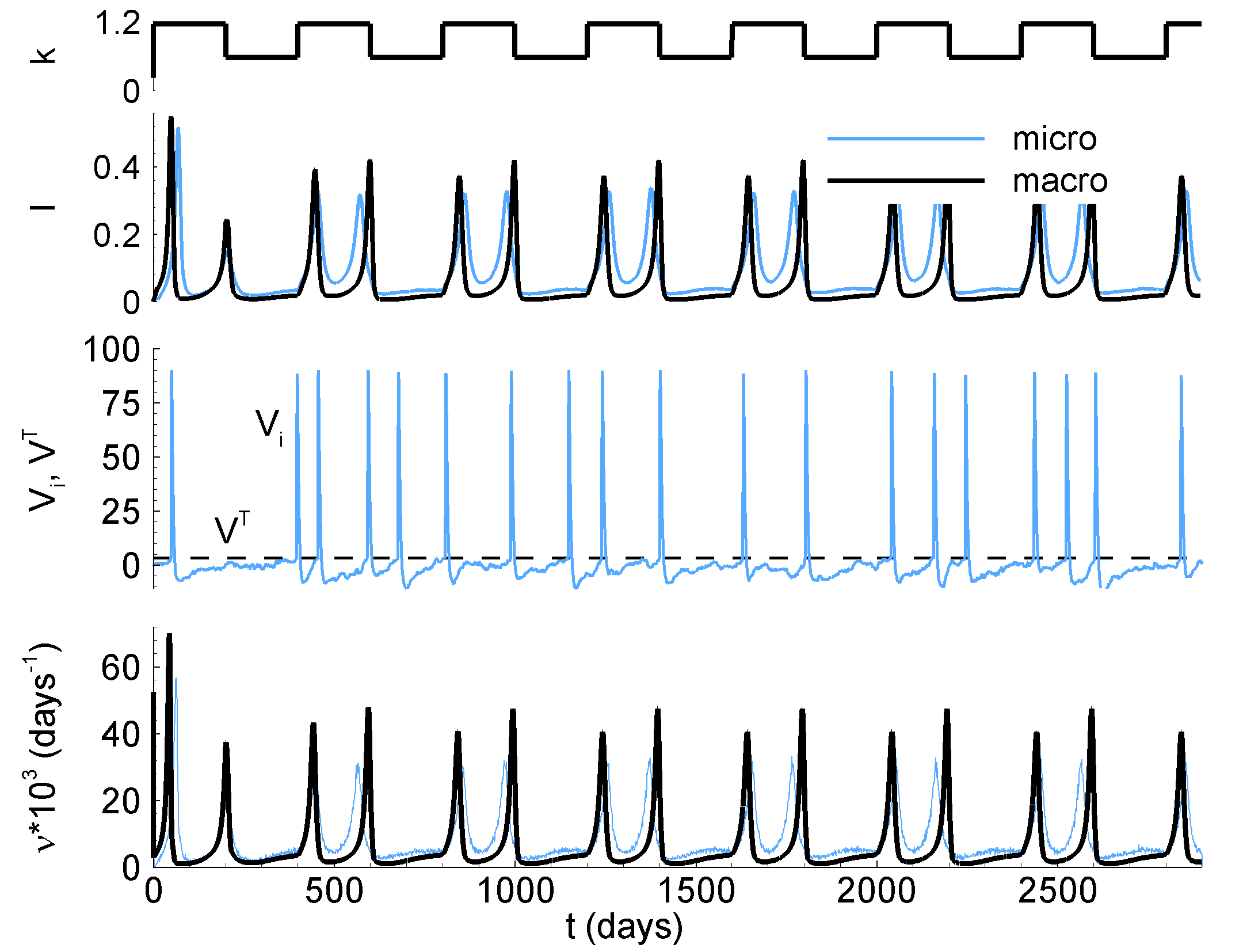}}
\caption{White noise. Simulation of the response to the rapid change of the rate of viral load $k$ from 3 to 1.2 and the noise amplitude $\sigma(t)$ from 0.5 to 2, at $t=t_0=0$. {\bf A}, top to bottom: $k(t)$, $\sigma(t)$, the fraction of potentially infected population $I(t)$, the state variable $V_i(t)$ of a representative individual, and the rate of new cases $\nu(t)$. 
For $I(t)$ and $\nu(t)$, the solutions obtained with both macro (black lines) and microscopic equations for $N=20000$ (blue lines) are shown. {\bf B}, the distribution in the $\ts$-space of the mean state variable $U$, the source term $\rho H$, and the density $\rho$, in 4 time moments. {\bf C}, Response to periodically changing $k$ mimicking change of seasons ($\VT=4$). {\bf D}, Response to periodically changing $k$ mimicking change of seasons ($\VT=3.7$). 
Parameters: 
$\tau=50$, $\tau_I=10$, $\tau_r=1$, $\tau_R=20$, $a=50$, $\VT=4$, and $\sigma=2\sqrt{\pi}$. }
\label{fig:U_nu}
\end{figure}

\paragraph{Model with white Gaussian noise} 
When simulating the system with the white Gaussian noise (Fig. \ref{fig:U_nu}), we observed a similar behavior. The onset of an epidemic is again modeled 
by a short pulse of the rate of new infections $\nu(t)$ of magnitude $I_0$ at time $t=0$ reflecting the appearance of infectious individuals. At the same time, the viral load noise jumps ($\sigma$ changes from $0.5$ to $2$). These conditions result in sustained oscillations of the rate of cases $\nu(t)$ and the fraction of potentially infected population $I(t)$. Again, the dynamics of a single individual state $V_i(t)$ (Fig. \ref{fig:U_nu}A, next to bottom panel) follows the epidemic waves of $\nu(t)$ and $I(t)$. When at $t=t_1=900$ the basic reproduction rate $k$ drops due to assumed anti-epidemic measures, the epidemic wave generation stops.

\paragraph{Mesoscopic model} 

\begin{figure}[h!]
\centerline{\includegraphics[scale=0.14]{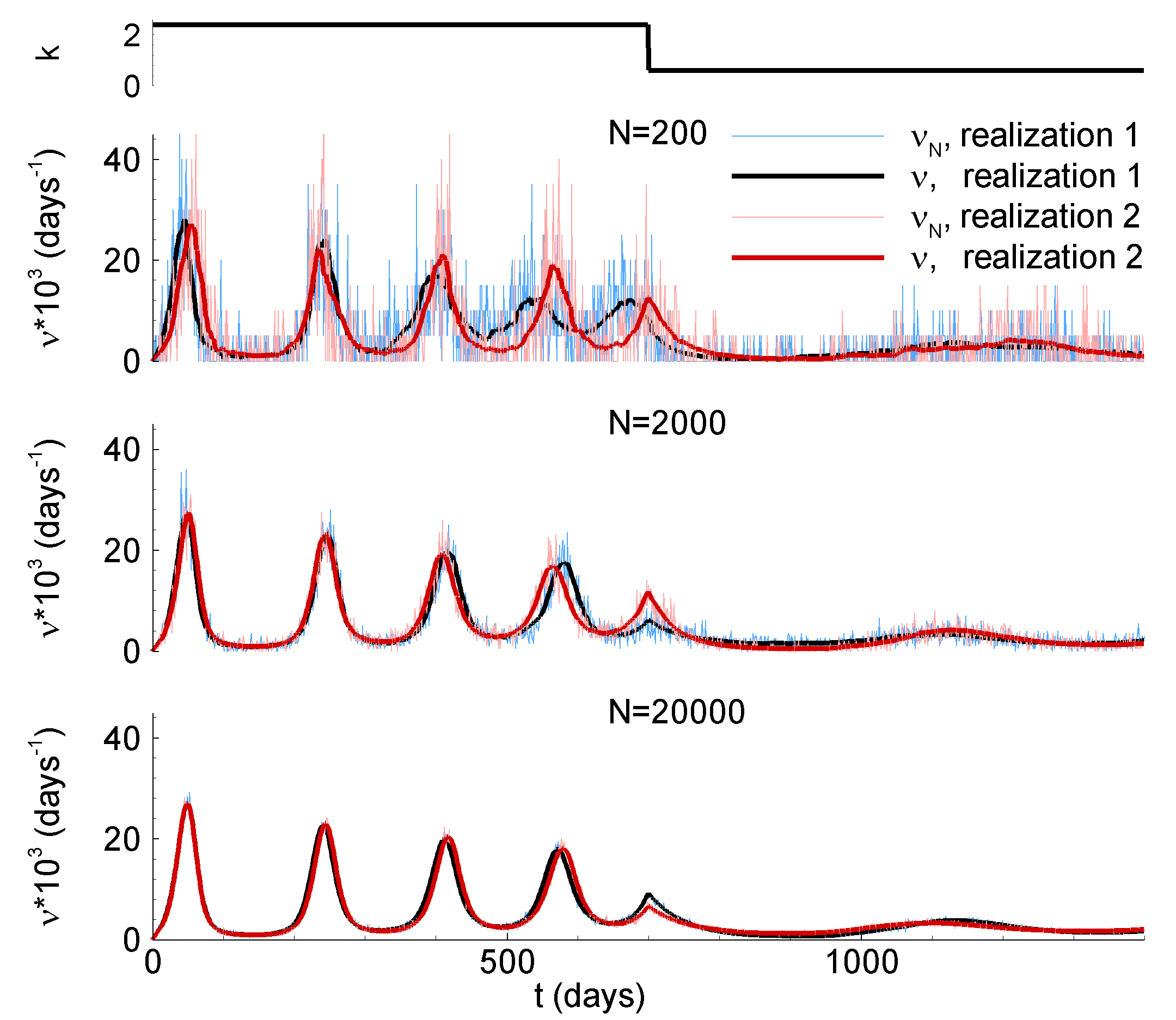}}
\caption{Mesoscopic model \eqref{Eq_escape} for different $N=200$, 2000, and 20000, and different realizations of noise. The parameters are as in Fig. \ref{fig:U_nu5}. The solutions with $N \xrightarrow{} \infty$ converge to the macroscopic model solution shown in Fig. \ref{fig:U_nu5}.}
\label{fig:Meso}
\end{figure}

In order to consider a finite number of individuals in a population, we apply the mesoscopic model described in section \ref{sec:meso}, based on Eqs.\eqref{EqMeso1}-\eqref{EqMeso7}. The solutions are different for different $N$ (Fig.\ref{fig:Meso}). The solutions for small $N$ differ significantly for different realizations of the noise. The model in the limit $N \xrightarrow{} \infty$ is equivalent to the macroscopic model. Hence, the solutions for large $N$ converge for different realizations and approach the macroscopic model solution shown in Fig. \ref{fig:U_nu5}.

\section{Discussion}
\label{sec:discussion}
In the present work, we have proposed the description of an epidemic that stems from a microscopic model of an individual and results in the meso- and macroscopic models of the entire population. The microscopic evolution is described by the dynamics of two state variables for each individual, the interplay between the viral load and immune response $V_i$, and the time since the last infection $\ts_i$. The corresponding mesoscopic and macroscopic models are then given by deterministic and stochastic refractory-density equations, respectively. We have shown the consistency between the models. The presented simulations remind of epidemic ``waves'', for instance, the ones reported in \cite{Nickbakhsh2020}. One of our basic ideas to describe the epidemic in terms of the evolution of viral load depending on the time since the last infection is supported by the experimental data from \cite{JonBie21}, where the viral load was measured in patients with SARS-CoV-2 as a function of the time since infection. Our multi scale modeling framework allows us to take into account such microscopic parameters measured in individuals in order to constrain the parameters of the meso- or macro-scale models.

The approach and the model that we presented in this manuscript could naturally be extended and generalized. We conclude the manuscript with a few promising leads for future research, which we identified during the writing of this work.

\subsection{Treatment}
In simulations, we varied the individual rate of viral load  $k(t)$, which accounted for the seasonality. This factor can also reflect such social measures as wearing masks and isolation. 
In contrast, medical treatment of patients is a function of the time since infection $\ts$, and it changes the course of the disease. Naturally, this factor can be taken into account through modification of the shape of the function $D$ entering the equation for the interplay of viral load and immune response, Eq. \eqref{Eq1'}.

\subsection{Modelling infectiousness depending on infection age}
It has been shown that the infectiousness of SARS-CoV-2 depends on the time since the last infection \cite{JonBie21}. This can be easily modeled in our framework by generalizing Eq.~\eqref{eq:I}, Eq.~\eqref{Eq47'} and Eq.~\eqref{Eq47''} for the micro-, macro- and mesoscopic fractions of infected people to
\begin{equation}
    I(t)=\frac{1}{N}\int_{0}^t\kappa(t-s)\,dN_c(s),\quad I(t)=\int_{0}^t\kappa(t-s)\nu(s)\,ds\quad\text{and}\quad I(t)=\int_{0}^t\kappa(t-s)\nu_N(s)\,ds,
    \label{eq:I-generalized}
\end{equation}
respectively.
The kernel $\kappa:\mathbb{R}^+\to [0,1]$ describes the infectiousness depending on the infection course.
Our equations~\eqref{eq:I}, ~\eqref{Eq47'} and Eq.~\eqref{Eq47''} are recovered as the special case $\kappa(t^*)=\mathbbm{1}_{(0,\tau_I)}(t^*)$, allowing only two possible states -- infectious and non-infectious. Our infectious period $\tau_I=10~$days well correspond to the half-duration of cell culture infectivity measured in \cite{JonBie21} for SARS-CoV-2.

\subsection{Network of multiple populations}
Certain types of heterogeneity (\textit{e.g.} different social communities, age groups, etc.) and spatial structure (cities) can be treated by splitting a large heterogeneous population into smaller homogeneous subpopulations, again in analogy to RDE application in neuroscience, for instance, to simulate cortical neuronal populations \cite{ChiGra21}. Moreover, the methodology can be extended to interacting populations, such as in the scenario of epidemic transmission between countries, consideration of a network of communities, similar to the so-called multi-group approach (see e.g. \cite{adimy2023multigroup,ottaviano2023global}). Since the multi-group approach naturally operates with smaller-size subpopulations, the proposed mesoscopic approach could be applied in this case.

\subsection{Multiple internal variables}
The proposed derivation of the population model from that for an individual applies not only to the 1-D individual model described by a single ODE for the viral state $V_i$, but also to multidimensional cases, in which one may be interested in including additional characteristics of individuals and/or of populations, and keep track of these as the disease spreads. In this way, additional internal variables can be introduced, describing, for instance, activation and/or inactivation of some processes in the immune system. This concept resembles the consideration of Hodgkin-Huxley-like neurons in neuroscience  \cite{ChizhovGraham2007}. If an individual is characterized by $n$ ODEs, the population model comprises $n+1$ partial differential equations (PDEs) for the state variable and the density. These equations remain one-dimensional,  with time since the infection $\ts$ is the sole independent variable, besides $t$. Consequently, this approach maintains computational efficiency even for multi-dimensional individuals. Leveraging this advantage, more intricate models can be developed.

In comparison with the above-mentioned kinetic theory \cite{loy2021boltzmann} that results in the Fokker-Planck equation describing the evolution of the population in the phase space of a state variable, the proposed approach considers the evolution in the phase space of the time since infection. Once the microscopic models are identical, these two macroscopic approaches would result in very similar numerical simulations, at least in the case of white Gaussian noise \cite{SchwalgerChizhov2019}. In the more general case of non-scalar state variables, the Fokker-Planck equation becomes multidimensional, whereas the transport equations of the refractory density approach remain 1-D PDEs that are effectively solvable.

\subsection{Age-dependence}
The epidemics to a different extent affect people of different ages \cite{JonBie21}. Our approach allows the consideration of the population distributed by age $\hat{a}$ via an age-dependent hazard function $H(V,t^*,\hat{a})$. In this case, the main variables have to be parameterized by $\hat{a}$ as $\rho=\rho(t,\ts, \hat{a})$  and  $U=U(t,\ts, \hat{a})$. Assuming that the hazard function changes much slower with age $\hat{a}$ than with the time since the last infection (i.e. $|\partial H/\partial \hat{a}|\ll |\partial H/\partial t^*|$), the age $\hat{a}$ can be treated quasi-statically as a parameter and not as an independent dynamical variable. Under this plausible assumption, the equations of the macroscopic model remain to be 1-D transport equations. However, the fraction of infected individuals $I(t)$ would be an integral over $\hat{a}$. This parametrization would allow the consideration of different disease time courses, etc. The infectivity $\kappa(\ts, \hat{a})$ can also be dependent on $\hat{a}$ because usually people interact stronger within their age groups.  

\subsection{Asymptomatic spread} 
We remark that, for our model, we are interested in the ability of an individual to spread the disease, rather than showing symptoms; hence, the infected and infectious population might be generalized to include asymptomatic individuals, thus allowing to adapt our construction to more complex compartmental models such as the ones presented in \cite{ansumali2020modelling,arino2020simple,huo2021estimating,ottaviano2022global,ottaviano2023global}. This may be done in multiple ways, depending on which characteristics of symptomatic and asymptomatic spread one wishes to capture in their model.

\subsection{Heterogeneous severity} 
It could be of biological interest to assume a heterogeneous setting, e.g. the severity $a$ in Eq.~\eqref{Eq21'} could vary among individuals depending on the vaccine status, stronger or weaker immune system, etc.; whereas we can assume that $\tau_r$ is approximately constant and depends only on the specific disease. However, this generalization would carry a non-negligible increase in the computational cost of our simulations, and make mean field approximations which we perform in section \ref{sec:population} and \ref{sec:meso} considerably more difficult. One potential strategy to deal with such heterogeneity would be to split the population into several, approximately homogeneous subpopulations and use a multi-population version of our model as described above. Importantly, in this scenario, we could take full advantage of the mesoscopic model because it remains valid for small sub-populations.
We leave such a generalization to heterogeneous systems as a promising outlook for future work.

\subsection{Complexity and parameters} 
Despite more mathematical complexity in comparison with the classical SEIRS models (PDEs versus ODEs), the proposed model has a similar number of parameters. The SEIRS model has 5 parameters ($\beta$, $\sigma$, $\gamma$, and $\xi$, in traditional notations, and the time scale). Instead, in the case of omitted explicit course of the recovery $D(t)$, the proposed model has also only 5 parameters: $\tau$, $\tau_I$, $k$, $\sigma$, and $\VT$.

\subsection{Inverse problem} 
The model may be used in future studies for solving reverse problems of finding the parameters from statistical data for a certain disease. For this purpose, one should have a sufficient set of experimental data that would reflect the effects of each of the parameters. For instance, the effect of $k$ can be studied with the effect of masks; the effect of $\tau_I$ with the effect of care; the effect of $\VT$ with the effect of provisional stimulation of the immune system of the entire population; the effect of $\sigma$ with a selection of more homogeneous subpopulation; the effect of $\tau$ with the selection of faster (or slower) recovering subpopulation, and so on.

\subsection{Limitations} 
In this first work on the topic, we made several simplifying assumptions. One of the most important is that
our model assumes a strictly conserved population: no removal, no mortality, closed community: no external influx, no out-flux. It would be of interest to consider, instead, a similar model in which the population is allowed to vary in time, for any of the reasons listed above.

\section*{Declaration of Competing Interest}

The authors have no competing interest to declare.

\section*{Declaration of equal contribution}

All authors contributed equally to this work.

\section*{Code availability} 
The code used to numerically solve the models presented in this manuscript is available at \\\url{https://github.com/schwalger/refracdens_epidem}.

\section*{Acknowledgements} Mattia Sensi was partially supported by the Italian Ministry for University and Research (MUR) through the PRIN 2020 project ``Integrated Mathematical Approaches to Socio-Epidemiological Dynamics'' (No. 2020JLWP23, CUP: E15F21005420006).
\appendix
\section{Macroscopic vs microscopic models, white Gaussian \textit{vs.} escape noise}

\begin{figure}[t]
\centerline{\includegraphics[scale=0.14]{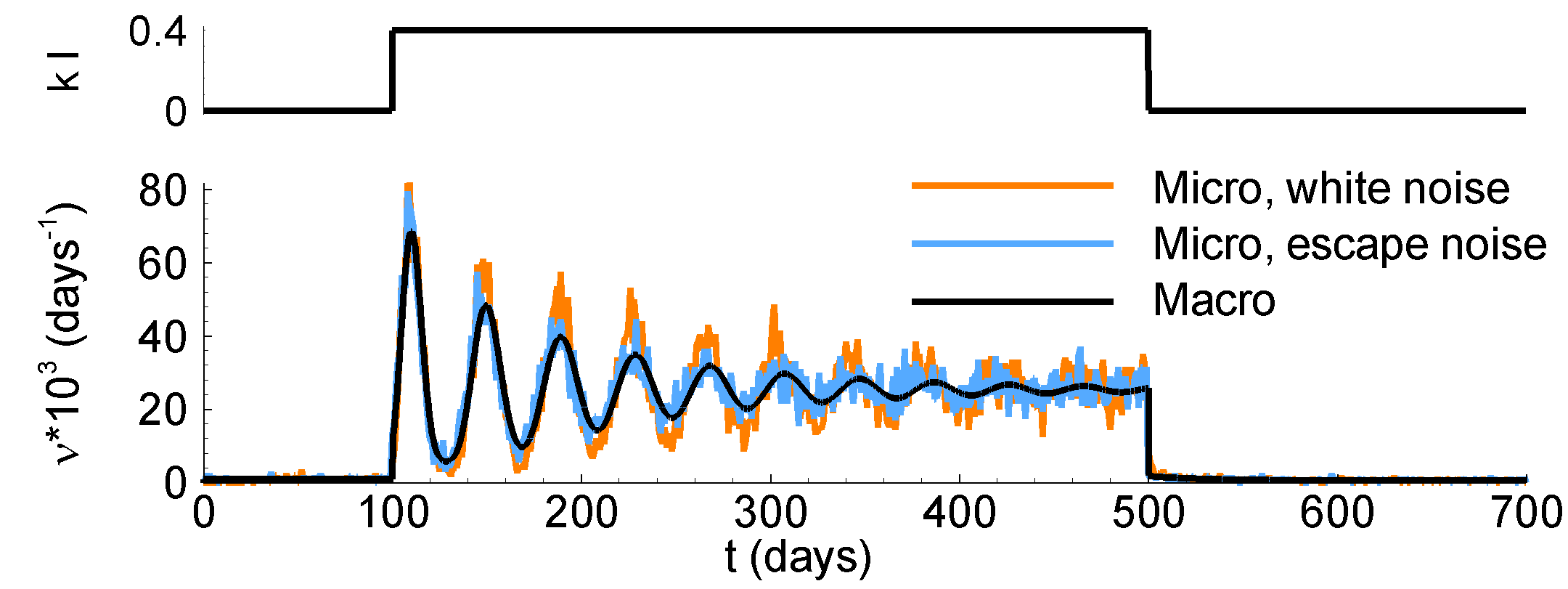} }
\caption{Solutions of a test with a stepwise term instead of $k(t)I(t)$ in the Eqs. \eqref{Eq1'} and \eqref{Eq47'}, obtained with the microscopic model ($N=2000$) with the white Gaussian noise and the escape noise described with Eq. \eqref{Eq8}, and the microscopic model. The value of $k(t)I(t)$ is 0 for $t\in [0,100)$ and $t >500$ and $0.4$ for $t \in [100,500]$.}
\label{fig:nu_WN_escape}
\end{figure}

To illustrate the agreement between microscopic and mesoscopic models as well as escape noise and Gaussian white noise, we consider a slightly simplified setup. Whereas the main model equation \eqref{Eq1'} includes the term $k(t) I(t)$, which is recurrently dependent on the activity of the entire population $\nu(t)$, here we consider a simplified problem with a step-wise input signal instead of the term $k(t) I(t)$. This case can be interpreted as a rough approximation of an epidemic problem, where the time course of the epidemic is assumed to be known and shaped as a step function. In this interpretation, we are interested in the probabilistic behavior of an individual, with $\nu(t)$ to be a probability for this individual to get ill. We simulate this response for the case of white noise and the escape noise in the form of Eq. \eqref{Eq8} (Fig.\ref{fig:nu_WN_escape}).

From the methodical point of view, Fig. \ref{fig:nu_WN_escape} shows the testing comparison of the macroscopic model with the microscopic models with white Gaussian and escape noise. The microscopic solutions converge to the macroscopic ones.

\bibliographystyle{plain}
\bibliography{biblio}

\begin{thebibliography}{10}

\bibitem{adimy2023multigroup}
M.~Adimy, A.~Chekroun, L.~Pujo-Menjouet, and M.~Sensi.
\newblock A multigroup approach to delayed prion production.
\newblock {\em Discrete and Continuous Dynamical Systems-B}, 2023.

\bibitem{alfaro2019analysis}
J.~A. Alfaro-Murillo, Z.~Feng, and J.~W. Glasser.
\newblock Analysis of an epidemiological model structured by
  time-since-last-infection.
\newblock {\em Journal of differential equations}, 267(10):5631--5661, 2019.

\bibitem{And88}
R.~M. Anderson.
\newblock The epidemiology of {HIV} infection: variable incubation plus
  infectious periods and heterogeneity in sexual activity.
\newblock {\em J. Roy. Stat. Soc.: Series A (Statistics in Society)},
  151(1):66--93, 1988.

\bibitem{ansumali2020modelling}
S.~Ansumali, S.~Kaushal, A.~Kumar, M.~K. Prakash, and M.~Vidyasagar.
\newblock Modelling a pandemic with asymptomatic patients, impact of lockdown
  and herd immunity, with applications to {SARS-CoV-2}.
\newblock {\em Annual reviews in control}, 50:432--447, 2020.

\bibitem{arino2020simple}
J.~Arino and S.~Portet.
\newblock A simple model for {COVID-19}.
\newblock {\em Infectious Disease Modelling}, 5:309--315, 2020.

\bibitem{brauer2009age}
F.~Brauer and J.~Watmough.
\newblock Age of infection epidemic models with heterogeneous mixing.
\newblock {\em Journal of biological dynamics}, 3(2-3):324--330, 2009.

\bibitem{ChizhovGraham2007}
A.~V. Chizhov and L.~J. Graham.
\newblock Population model of hippocampal pyramidal neurons, linking a
  refractory density approach to conductance-based neurons.
\newblock {\em Physical Review E}, 75(1):011924, 2007.

\bibitem{ChiGra08}
A.~V. Chizhov and L.~J. Graham.
\newblock Efficient evaluation of neuron populations receiving colored-noise
  current based on a refractory density method.
\newblock {\em Phys. Rev. E}, 77(1):011910, 2008.

\bibitem{ChiGra21}
A.~V. Chizhov and L.~J. Graham.
\newblock A strategy for mapping biophysical to abstract neuronal network
  models applied to primary visual cortex.
\newblock {\em PLoS Comput. Biol.}, 17(8):e1009007, 2021.

\bibitem{della2022sir}
R.~Della~Marca, N.~Loy, and A.~Tosin.
\newblock An {SIR--like kinetic model tracking individuals' viral load}.
\newblock {\em Networks and Heterogeneous Media}, 17(3):467--494, 2022.

\bibitem{DumHen24}
G.~Dumont, J.~Henry, and C.~O. Tarniceriu.
\newblock {Oscillations in a Fully Connected Network of Leaky
  Integrate-and-Fire Neurons with a Poisson Spiking Mechanism}.
\newblock {\em J. Nonlin. Sci}, 34(1):18, 2024.

\bibitem{DumTar24}
G.~Dumont and C.~O. Tarniceriu.
\newblock {Pattern Formation in a Spiking Neural-Field of Renewal Neurons},
  2024.

\bibitem{d2023optimal}
A.~d’Onofrio, M.~Iannelli, P.~Manfredi, and G.~Marinoschi.
\newblock Optimal epidemic control by social distancing and vaccination of an
  infection structured by time since infection: the {COVID-19} case study.
\newblock {\em SIAM Journal on Applied Mathematics}, pages S199--S224, 2023.

\bibitem{gerstner_spiking_2002}
W.~Gerstner and W.~M. Kistler.
\newblock {\em {Spiking {Neuron} {Models}: {Single} {Neurons}, {Populations},
  {Plasticity}}}.
\newblock Cambridge University Press, August 2002.
\newblock Google-Books-ID: Rs4oc7HfxIUC.

\bibitem{huo2021estimating}
X.~Huo, J.~Chen, and S.~Ruan.
\newblock Estimating asymptomatic, undetected and total cases for the
  {COVID-19} outbreak in {W}uhan: a mathematical modeling study.
\newblock {\em BMC Infectious Diseases}, 21(1):476, 2021.

\bibitem{JonBie21}
T.~C. Jones, G.~Biele, B.~Mühlemann, T.~Veith, J.~Schneider,
  J.~Beheim-Schwarzbach, T.~Bleicker, J.~Tesch, M.~L. Schmidt, L.~E. Sander,
  Fl. Kurth, P.~Menzel, R.~Schwarzer, M.~Zuchowski, J.~Hofmann, A.~Krumbholz,
  A.~Stein, A.~Edelmann, V.~M. Corman, and C.~Drosten.
\newblock Estimating infectiousness throughout {SARS-CoV-2} infection course.
\newblock {\em Science}, 373(6551):eabi5273, 2021.

\bibitem{kermack1927contribution}
W.~O. Kermack and A.~G. McKendrick.
\newblock A contribution to the mathematical theory of epidemics.
\newblock {\em Proceedings of the royal society of london. Series A, Containing
  papers of a mathematical and physical character}, 115(772):700--721, 1927.

\bibitem{li2019mathematical}
L.~L. Li, C.~P. Ferreira, and B.~Ainseba.
\newblock Mathematical analysis of an age structured problem modelling
  phenotypic plasticity in mosquito behaviour.
\newblock {\em Nonlinear Analysis: Real World Applications}, 48:410--423, 2019.

\bibitem{li2002global}
M.~Y. Li and L.~Wang.
\newblock Global stability in some {SEIR} epidemic models.
\newblock In {\em Mathematical approaches for emerging and reemerging
  infectious diseases: models, methods, and theory}, pages 295--311. Springer,
  2002.

\bibitem{loy2021boltzmann}
N.~Loy and A.~Tosin.
\newblock Boltzmann-type equations for multi-agent systems with label
  switching.
\newblock {\em Kinetic and Related Models}, 14(5):867--894, 2021.

\bibitem{Nickbakhsh2020}
S.~Nickbakhsh, A.~Ho, D.~F.~P. Marques, J.~McMenamin, R.~N. Gunson, and P.~R.
  Murcia.
\newblock Epidemiology of seasonal coronaviruses: establishing the context for
  the emergence of coronavirus disease 2019.
\newblock {\em The Journal of infectious diseases}, 222(1):17--25, 2020.

\bibitem{ottaviano2022global}
S.~Ottaviano, M.~Sensi, and S.~Sottile.
\newblock Global stability of {SAIRS} epidemic models.
\newblock {\em Nonlinear Analysis: Real World Applications}, 65:103501, 2022.

\bibitem{ottaviano2023global}
S.~Ottaviano, M.~Sensi, and S.~Sottile.
\newblock Global stability of multi-group {SAIRS} epidemic models.
\newblock {\em Mathematical Methods in the Applied Sciences},
  46(13):14045--14071, 2023.

\bibitem{PakPer09}
K.~Pakdaman, B.~Perthame, and D.~Salort.
\newblock Dynamics of a structured neuron population.
\newblock {\em Nonlinearity}, 23(1):55--75, 2009.

\bibitem{Schmutz2023}
V.~Schmutz, E.~L\"ocherbach, and T.~Schwalger.
\newblock {On a Finite-Size Neuronal Population Equation}.
\newblock {\em SIAM J. Applied Dynamical Systems}, 22(2):996--1029, 2023.

\bibitem{Sch21}
T.~Schwalger.
\newblock {Mapping Input Noise to Escape Noise in Integrate-and-fire neurons: A
  Level-Crossing Approach}.
\newblock {\em Biol. Cybern.}, 115:539--562, 2021.

\bibitem{SchwalgerChizhov2019}
T.~Schwalger and A.~V. Chizhov.
\newblock Mind the last spike—firing rate models for mesoscopic populations
  of spiking neurons.
\newblock {\em Current opinion in neurobiology}, 58:155--166, 2019.

\bibitem{Schwalger2017}
T.~Schwalger, M.~Deger, and W.~Gerstner.
\newblock Towards a theory of cortical columns: {F}rom spiking neurons to
  interacting neural populations of finite size.
\newblock {\em PLoS Comput Biol}, 13(4):e1005507, 2017.

\bibitem{Tarn20}
C.~O. Tarniceriu.
\newblock {Age-structure in neuronal models}.
\newblock {\em Annals of the Alexandru Ioan Cuza University-Mathematics},
  66(2):385--396, 2020.

\end{thebibliography}

\end{document}